\RequirePackage{lineno}

\documentclass[a4paper,11pt]{article}
\pdfoutput=1 %
\usepackage{jcappub} 

\usepackage[utf8]{inputenc}

\usepackage{color}
\usepackage{xcolor}
\usepackage{url}

\usepackage{graphicx}

\usepackage{graphics}
\usepackage{natbib}
\graphicspath{{./}}

\usepackage[normalem]{ulem}

\usepackage{soul}

\usepackage{supertabular}
\usepackage{tabularx}

\def\unit#1{\hbox{$\,{\rm #1}$}}
\def\degrees{\hbox{${}^\circ$}}

\newcommand{\vv}[1]{``{#1}''}

\title{Constraints on dark matter scattering with long lived mediators from observations of the Sun with the Fermi Large Area Telescope}

\author[a,1]{D.~Serini,\note{Corresponding authors.}}
\author[a,b,1]{F.~Loparco}
\author[a,1]{M.~N.~Mazziotta}
\author[a,b]{S.~De~Gaetano}
\author[a,b]{L.~Di~Venere}
\author[a]{F.~Gargano}
\author[a,b]{L.~Lorusso}
\author[a,b]{G.~Panzarini}
\author[a,b]{R.~Pillera}

\affiliation[a]{Istituto Nazionale di Fisica Nucleare, Sezione di Bari, via Orabona 4, I-70126 Bari, Italy}
\affiliation[b]{Dipartimento di Fisica ``M. Merlin" dell'Universit\`a e del Politecnico di Bari, via Amendola 173, I-70126 Bari, Italy}

\emailAdd{davide.serini@ba.infn.it}

\emailAdd{francesco.loparco@ba.infn.it}

\emailAdd{mazziotta@ba.infn.it}

\date{\today}

\abstract{
The Sun represents a promising target for indirect dark matter searches, as dark matter particles from the Galactic halo can be gravitationally trapped in its core or in external orbits, and their annihilations can lead to final states with standard model particles that are able to reach the Earth. In this work we have considered a scenario in which dark matter particles can annihilate into pairs of long-lived mediators, which in turn can escape from the Sun and decay into pairs of gamma rays or into the $b\bar{b}$, $\tau^{+}\tau^{-}$, $\mu^{+}\mu^{-}$ channels, with the production of gamma rays in the final states. All these processes are expected to yield an excess in the energy spectrum of gamma rays towards the Sun. We have therefore analyzed the data collected by the Fermi Large Area Telescope during its first 13.5 years of operation, searching for possible excesses in the solar gamma-ray spectrum. Since no statistically significant excess is found, we have set constraints on the dark matter-nucleon scattering cross sections in both the spin-dependent and spin-independent cases. For all the mediator decay channels explored and for dark matter masses between a few $\unit{GeV/c^2}$ and $1\unit{TeV/c^2}$, we have found that the upper limits on the spin-dependent and spin-independent cross sections are in the ranges from $10^{-45}$ to $10^{-39}$ $\unit{cm^{2}}$ and from $10^{-47}$ up to $10^{-42}$ $\unit{cm^{2}}$, respectively. 
}

\keywords{Sun, Gamma-Rays, Dark Matter}

\begin{document}

\maketitle

\flushbottom

\section{Introduction}
\label{sec:intro}

The Sun is among the potential targets for indirect dark matter (DM) searches. DM particles from the Milky Way halo can undergo elastic interactions with the solar nuclei, in which they progressively lose energy until they are trapped by the solar gravitational field and sunk into the core of the Sun. DM particles captured inside the Sun can then self-annihilate into standard model (SM) particles that, with the exception of high-energy neutrinos, are likely re-absorbed in the interior of the Sun and cannot reach the Earth. However, in recent works~\cite{Pospelov:2007mp,ArkaniHamed:2008qn,Schuster:2009fc,Leane:2017vag,Schuster:2009au,Bell:2011sn,Arina:2017sng} an alternative scenario was proposed, in which pairs of DM particles $\chi$ annihilate into pairs of long-lived mediators $\phi$, which can escape from the Sun and then decay, yielding SM particles in the final states, such as gamma rays or electron-positron pairs, that can reach the Earth and be detected.

In our previous works~\cite{mazziotta2020search,cuoco2020search} we analyzed the data collected by the Large Area Telescope (LAT) onboard the {\it Fermi} Gamma-ray Space Telescope mission~\cite{Fermi-LAT:2009ihh} during its first 10 years of operation to search for possible signals from solar DM in the gamma-ray and in the electron-positron channels, and we were able to derive constraints on the DM-nucleon scattering cross sections. These constraints were obtained under the assumption that gamma rays or electron-positron pairs are directly produced in the mediator decays, i.e. in the processes $\phi \rightarrow \gamma \gamma$ or $\phi \rightarrow e^{+}e^{-}$. Under these assumptions, the resulting spectra of gamma rays and of electrons-positrons from DM are expected to exhibit a box-like feature, with the upper edge of the box corresponding to the mass $m_{\chi}$ of the candidate DM particle~\cite{Ibarra:2012dw}.

In this work we have updated the dataset of Ref.~\cite{mazziotta2020search}, including the data collected by the LAT in its first 13.5 years of operation. In addition, we have implemented a more general analysis approach, which can be also applied to search for possible solar DM signals from WIMP annihilations into long-lived mediators decaying into several different channels with gamma rays in the final states. In all these scenarios, the possible solar DM gamma-ray signals extend over a wide energy range, and the evaluation of the energy spectra requires a dedicated simulation, which is discussed in Sec.~\ref{sec:medchannelstheory}. 

We have studied the spectra of gamma rays from a small Region of Interest (RoI) centered on the Sun and we have evaluated the possible signal counts from DM following two different approaches discussed in Sec.~\ref{sec:analysis}. The constraints on the DM signal intensities have been evaluated by requiring that the expected counts from DM annihilations in any channel do not exceed the possible signal counts.

\section{Gamma rays from solar dark matter}
\label{sec:medchannelstheory}

In this work we are considering a physics scenario with pairs of DM particles $\chi$ annihilating inside the Sun into pairs of long-lived mediators $\phi$, which can escape from the Sun and then decay into channels with gamma rays in the final states. These photons can be produced in the decay channel $\phi \rightarrow \gamma \gamma$ or as final state radiation (FSR) through the decay channels $\phi \rightarrow b\bar{b}, \tau^{+}\tau^{-}, \mu^{+}\mu^{-}, \ldots$ 

In the case of the $\phi \rightarrow \gamma \gamma$ decays, in Ref.~\cite{mazziotta2020search} we have shown that, for a light mediator (i.e. $m_{\phi} \ll m_{\chi}$), the gamma-ray spectrum at Earth can be well modeled as a simple box, with its upper edge at an energy corresponding to the mass of the DM particle $m_{\chi}$. Similary, in the case of FSR, the gamma-ray spectrum extends over an energy range up to $m_{\chi}$. In all these cases we have evaluated the expected gamma-ray fluxes at Earth using the \texttt{WimpSim}\footnote{\url{http://wimpsim.astroparticle.se/}} version 5.0 Monte Carlo simulation code~\cite{Niblaeus_2019}. 

The DM gamma-ray flux expected at Earth is given by:

\begin{equation}
      \Phi_{\text{DM}}(E;m_\chi,m_\phi,\sigma,L) = 
      \Gamma_{\text{cap}}(m_\chi, \sigma) \cdot 
      \varphi(E; m_\chi, m_\phi, L) 
\label{eq:generalphidm}
\end{equation}
where $E$ is the gamma-ray energy, $m_{\chi}$ and $m_{\phi}$ are the masses of the DM particle $\chi$ and of the mediator $\phi$ respectively, $\sigma$ is the elastic DM-nucleon scattering cross section and $L$ is the boosted mediator decay length. The term $\Gamma_{\text{cap}}(m_\chi, \sigma)$ in the r.h.s. of the previous equation is the DM capture rate in the Sun, which depends on  $m_{\chi}$ and $\sigma$, while the term $\varphi(E; m_\chi, m_\phi, L)$ is the flux at Earth of gamma rays per DM annihilation, which depends on the masses $m_{\chi}$ and $m_{\phi}$ and on the mediator decay length $L$. This term can be expressed as: 

\begin{equation}
\varphi(E; m_\chi, m_\phi, L)  =    
\frac{1}{4\pi D^2} Y(E;m_\chi,m_\phi,L)
\label{eq:phidm2}
\end{equation}
where $D=1 \unit{AU}$ is the Sun-Earth distance and $Y(E;m_\chi,m_\phi,L)$ is the gamma-ray yield per DM annihilation, which depends only on the kinematics of the mediator decay\footnote{The mass $m_\chi$ of the parent DM particle sets the energy of the mediator in the lab frame, since the annihilation process $\chi \chi \rightarrow \phi \phi$ is assumed to happen with both DM particles nearly at rest.}. Eq.~\ref{eq:generalphidm} has been written under the assumption of equilibrium between the DM capture and annihilation processes~\cite{cuoco2020search}. 

The DM capture rate is evaluated using the \texttt{DARKSUSY} code version \texttt{6.1.0}~\cite{Gondolo:2004sc,Bringmann:2018lay,darksusyweb}, under the assumptions of the default settings, with a local DM density $\rho_{\odot}=0.3\unit{GeV/cm^{3}}$, a Maxwellian velocity distribution with average velocity $v_{\odot}=20 \unit{km/s}$ and a velocity dispersion $v_{rms}=270 \unit{km/s}$. The calculation has been performed for both the spin-dependent and spin-independent cases. We have evaluated the capture rate using for the reference cross section $\sigma_{0}=10^{-40} \unit{cm^{2}}$. The capture rates corresponding to different cross sections can be easily obtained, since $\Gamma_{\text{cap}}$ is proportional to $\sigma$. Full details on the  evaluation of the capture rates are given in ref.~\cite{cuoco2020search}.

As mentioned above, to evaluate the gamma-ray yields, we have used the \texttt{WimpSim} code, that can simulate the production of neutrinos, gamma rays and other particles from DM annihilations in the Sun and in the Earth. The DM annihilations are simulated with the event generator \texttt{Pythia-6.4.26}\footnote{\url{http://home.thep.lu.se/~torbjorn/Pythia.html}}~\cite{sjostrand2015introduction}, which is also used to simulate the subsequent decay chains of the annihilation products. In our simulations we used the \texttt{med\_dec} module of \texttt{WimpSim}, which was specifically developed for describing the propagation and the decays of long-lived mediators. The mediators propagate out from the solar core, travelling along straight lines. The distance travelled before the decay is sampled from an exponential distribution, with an average boosted decay length $L$ set by the user. 

\begin{figure*}[!t]
\centering
    \includegraphics[width=\textwidth]{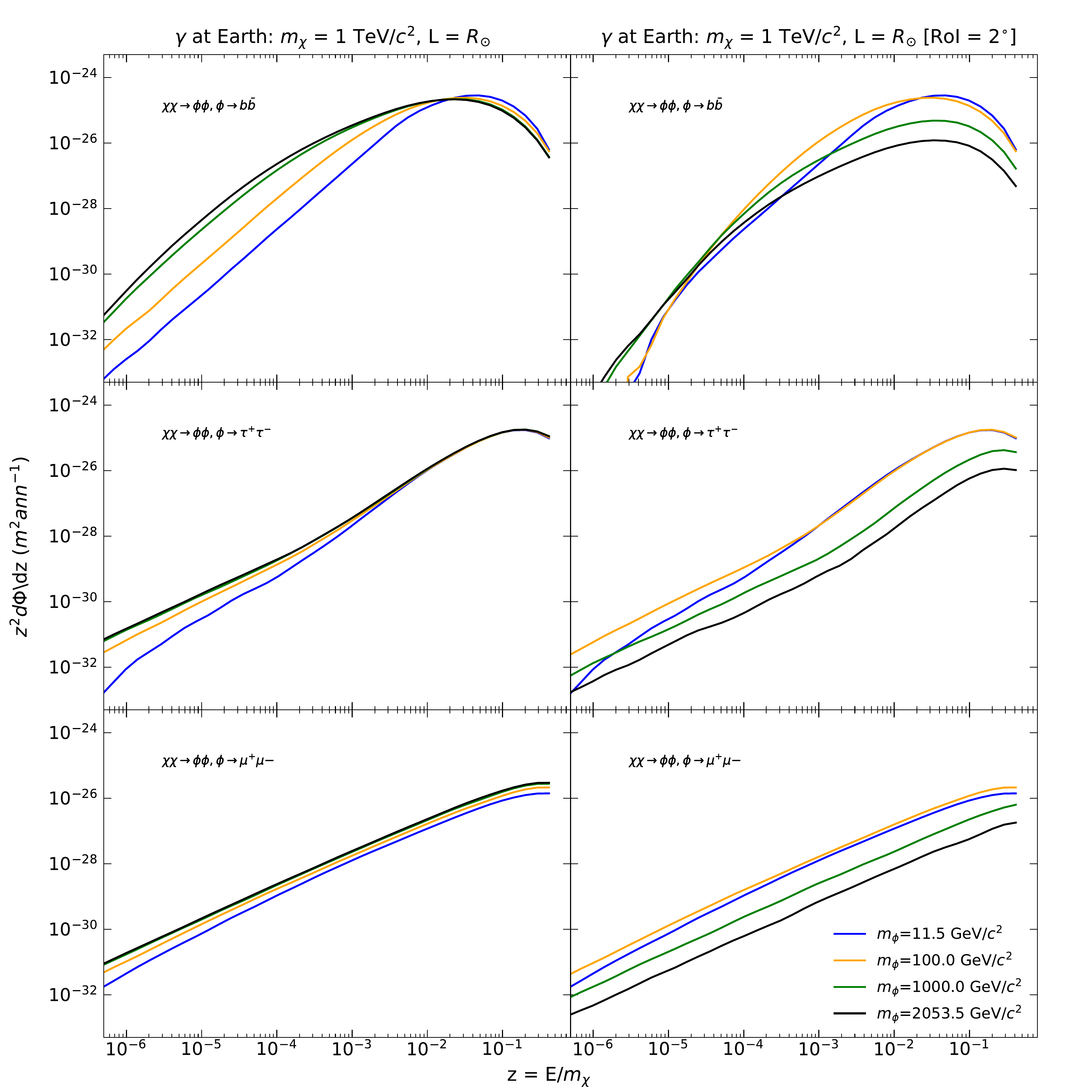}
    \caption{Expected gamma-ray fluxes at Earth per DM annihilation, evaluated with \texttt{med\_dec} for a $1 \unit{TeV}$ DM particle and different mediator masses, with a boosted decay length $L=R_{\odot}$. The plots refer to the decay channels $\phi \rightarrow b\bar{b}$ (top panels), $\phi \rightarrow \tau^{+} \tau^{-}$ (middle panels) and $\phi \rightarrow \mu^{+} \mu^{-}$ (bottom panels). The plots on the left panels show the simulation results obtained without any selection on the arrival directions of gamma rays at Earth, while those on the right panels show the results obtained when considering only photons with arrival directions lying within $2\degrees$ from the Sun. A total of $10^5$ events have been simulated for each configuration. The fluxes are evaluated in terms of the variable $z=E/m_{\chi}$.}.
 \label{fig:gammafluxes}
\end{figure*}

In this work we have studied several mediator decay channels, performing a dedicated simulation campaign, in which we have scanned a wide range of possible values of the masses $m_{\chi}$ and $m_{\phi}$. For each pair of values $(m_{\chi},m_{\phi})$, a set of $10^{5}$ events have been simulated. In our simulations we used \texttt{med\_dec} with the assumption $L=R_{\odot}$; the gamma-ray flux for different values of $L$ can be evaluated by taking into account the survival probability of the mediator in its journey from the Sun to the Earth~\cite{cuoco2020search}. After each simulation run, summary tables are generated with the phenomenological description of the events. In addition, event files are also generated, which contain the relevant information about the gamma rays produced in each event. The information stored in these files is used to evaluate the differential flux at Earth of gamma rays.

In Fig.~\ref{fig:gammafluxes} the outputs of some simulation runs are shown. The plots show the expected gamma-ray fluxes per DM annihilation from $1\unit{TeV}$ DM particles annihilating into mediators of different masses, which in turn can decay into the $b\bar{b}$, $\tau^{+} \tau^{-}$ and  $\mu^{+}\mu^{-}$ channels. The gamma-ray fluxes at Earth have been evaluated either without any selection on the arrival directions of gamma rays or selecting only those photons with arrival directions lying within $2\degrees$ from the Sun. This choice reflects the data selection implemented in the analysis (see Sec.~\ref{sec:analysis}). As can be seen from the examples shown in Fig.~\ref{fig:gammafluxes}, this selection reduces the intensity of the possible DM signal at low energies, i.e. in the region where $E/m_{\chi}<10^{-3}$. This is due to the fact that the angular dispersion of photons is $\sim 1/\gamma_{\phi}$, where $\gamma_{\phi}$ is the Lorentz factor of the mediator. Assuming that DM particles annihilate at rest, $\gamma_{\phi}=m_{\chi}/m_{\phi}$ and, therefore, the heavier the mediator, the higher will be the angular dispersion and the lower will be the fraction of photons within a cone of $2\degrees$ aperture from the Sun.

\section{Data selection and analysis}
\label{sec:analysis}

As mentioned above, in this work we have analyzed a data sample collected by the Fermi LAT during its first 13.5 years of operation. We have implemented an event selection based on our previous work~\cite{mazziotta2020search} and we have developed a novel analysis procedure, which allows us to constrain both scenarios with direct and indirect gamma-ray production in the mediator decays. 

\subsection{Data selection}

We have analyzed a dataset consisting of \texttt{P8R3\_CLEAN} photon events~\cite{Atwood:2013rka} collected by the Fermi LAT during its first 13.5 years of operation in the energy range above $100\unit{MeV}$. 

As in ref.~\cite{mazziotta2020search}, we have selected the data from two different RoIs, that we indicate as \vv{on} and \vv{off} respectively, corresponding to cones of $2\degrees$ angular radius centered on the Sun and on the anti-Sun directions. The direction of the anti-Sun is defined as the direction of the Sun with a forward/backward time offset of 6 months. This choice ensures that the angular separation between the Sun and the anti-Sun is always close to $180\degrees$ and that, during the LAT data-taking, the anti-Sun follows the same path in the sky as the Sun. The off region is therefore an optimal control region to take into account any possible systematic uncertainties.

The \vv{good time intervals} (GTIs) chosen for the analysis are those periods when the LAT was taking data in its nominal science operation configuration and was outside the South Atlantic Anomaly (SAA). Data taken during solar flares were also discarded. A maximum angular separation of $90\degrees$ was allowed between the direction of the Sun (anti-Sun) and the zenith, to avoid contamination from photons produced in the cosmic-ray interactions with the upper layers of the Earth's atmosphere. We also selected GTIs in which the angular separation between the Sun (anti-Sun) and the LAT z-axis was less than $70.5\degrees$. Finally, we discarded those time intervals when the Galactic latitude of the Sun (anti-Sun) was $|b|<7\degrees$, those when the angular separation of the Sun (anti-Sun) from the Moon was less than $7\degrees$ and those when the angular separation of the Sun (anti-Sun) from any bright\footnote{Here we define as \vv{bright} a source whose gamma-ray flux above $100 \unit{MeV}$ is larger than $4\times 10^{-7}$ \unit{photons~cm^{-2}s^{-1}}~\cite{mazziotta2020search}
. } source in the 4FGL Fermi LAT source catalog~\cite{Fermi-LAT:2019yla} was less than $7\degrees$.    

The observed gamma-ray energy distributions after the event selection in the on and off regions are similar to those shown in Fig.~2 of Ref.~\cite{mazziotta2020search}. Since the exposures of the two regions are nearly equal, the excess counts in the on region can be assumed to originate from the steady solar gamma-ray emission and from a possible DM signal. 

\subsection{Data analysis}

The Sun is a bright source of gamma rays, due to the interactions of cosmic rays (CRs) with the solar environment~\cite{Abdo:2011xn,Tang:2018wqp,Seckel:1991ffa,Orlando:2006zs,Orlando:2008uk}. The gamma-ray emission from the Sun includes a contribution from the disk, mainly due to the hadronic cascades produced in the interactions of CR nuclei with the solar atmosphere, and a diffuse contribution, due to the inverse Compton scatterings of CR electrons and positrons with the optical-UV solar photons in the heliosphere. These mechanisms yield a steady gamma-ray emission, with a continuous energy spectrum extending beyond the $\unit{TeV}$ region. Any possible DM signal will therefore appear as an excess over the steady gamma-ray emission spectrum. 

Modeling the steady solar gamma-ray emission is not straightforward, because of the complex and time-dependent structure of the solar magnetic field, which affects the trajectories of charged CRs in the solar environment and consequently also their interactions with the gases in the solar atmosphere. Although several attempts were performed in the past following different approaches~\cite{Seckel:1991ffa,Orlando:2006zs,Orlando:2008uk,Moskalenko:2006ta,mazziotta2020cosmic}, none of these models is able to accurately reproduce the gamma-ray flux measured by the Fermi LAT. 

For this reason we developed two approaches for the analysis of the LAT data which do not require any template model for the standard solar emission. In the first approach we assume that all the excess counts in the on region with respect to the off region entirely originate from a DM signal. Due to this assumption, the upper limits on the DM-nucleon cross section obtained with this approach will be clearly overestimated, since the steady solar emission is neglected. Conversely, in the second approach we assume that the excess counts in the on region originate from the steady solar emission, and gamma rays from DM annihilations are responsible only for possible fluctuations of these excess counts. The constraints on the DM-nucleon cross sections obtained with this approach therefore will be stronger than those obtained from the other approach. Hereafter we will indicate the two analysis approaches as \vv{conservative} and \vv{optimistic} respectively, with the conservative approach yielding weaker constraints than the optimistic one. 

The expected photon counts from DM annihilations in the on region are given by:
\begin{equation}
\mu_{DM}(E_{o})  =  t_{on} \int dE_{\text{t}} \mathcal{R}_{on}(E_{\text{o}}|E_{\text{t}})  
\Phi_{DM}(E_{\text{t}};m_{\chi},m_{\phi},\sigma,L) 
\label{eq:mus} 
\end{equation}
where $E_{\text{o}}$ is the reconstructed (observed) photon energy, $E_{\text{t}}$ is the true photon energy, $t_{on}$ is the integrated livetime of the on region, $\mathcal{R}_{\text{on}}(E_{\text{o}}|E_{\text{t}})$ is the instrument response matrix incorporating the effective area, the angular resolution and the energy resolution of the LAT.

The instrument response matrices are evaluated from the Monte Carlo simulations of the LAT, by selecting photon events with angular separations between the reconstructed and the true photon directions less than the angular radius of the RoI, i.e. $2\degrees$, and  taking into account the livetime distributions as a function of the off-axis angle in the instrument frame~\cite{Mazziotta:2009rd,Loparco:2009by}.

For each decay channel and for each pair of masses $(m_{\chi},m_{\phi})$, we have modeled the DM gamma-ray flux as:

\begin{equation}
\Phi_{DM}(E) = k_{DM} \Phi_{DM,0}(E;m_\chi,m_\phi,\sigma_{0},R_\odot) \label{eq:DMyieldFit}
\end{equation}
where the flux $\Phi_{DM,0}(E;m_\chi,m_\phi,\sigma_{0},R_\odot)$ is evaluated as discussed in Sec.~\ref{sec:medchannelstheory}, using a reference cross section value $\sigma_{0}=10^{-40}\unit{cm^2}$ and assuming a mediator decay length $L=R_{\odot}$. The reference fluxes have been evaluated for a set of DM masses and mediator masses, with the constraints $m_{\chi}>m_{\phi}$ and $m_{\phi}>2m_{X}$, where \text{X} is the daughter particle produced in the decays of the mediator $\phi$ ($X=\tau,b,\mu\ldots$). We have explored a range of DM masses up to $10\unit{TeV/c^2}$. The DM and mediator mass intervals have both been divided into 16 bins per decade, equally spaced on a logarithmic scale. 

\begin{figure*}[t]
\centering
    \includegraphics[width=0.98\columnwidth]{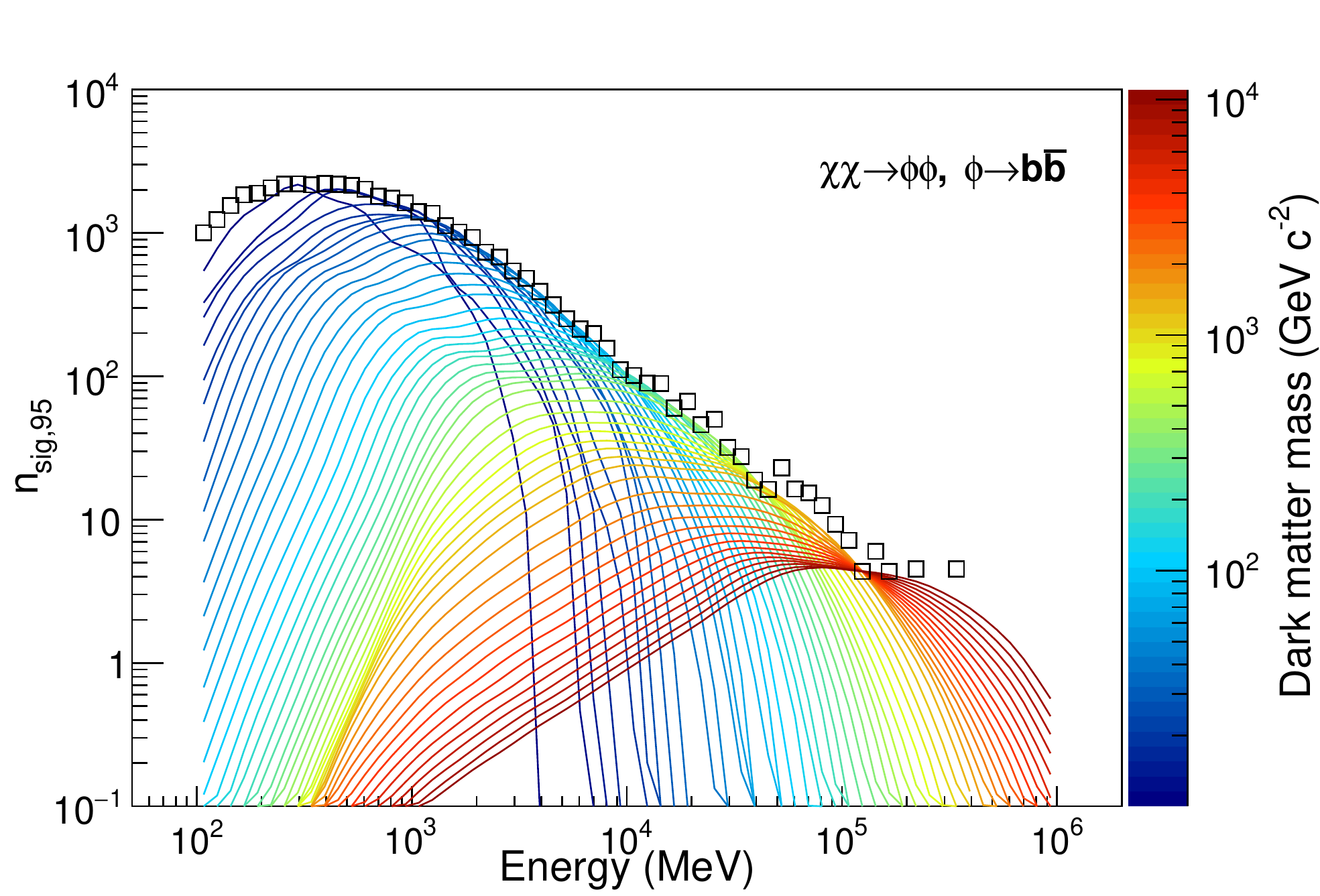}
    \caption{Evaluation of the upper limits at $95\%$ CL on the DM gamma-ray fluxes with the conservative approach. The plot refers to the mediator decay channel $\phi \rightarrow b\bar{b}$ with $m_{\phi}=11.5\unit{GeVc^{-2}}$. The open squares indicate the upper limits at $95\%$ CL on the signal counts $n_{sig,95}(E_{o})$, evaluated with the Bayesian method. Each colored line corresponds to the upper limit on the DM gamma-ray counts evaluated for a given value of the DM mass $m_{\chi}$. The values of $m_{\chi}$ are indicated on the color scale on the right. }
 \label{fig:calclimits}
\end{figure*}

The normalization $k_{DM}$ is then left as a free parameter. In the conservative approach it is evaluated by imposing that the counts from DM annihilations $\mu_{DM}(E_{o})$ do not exceed the upper limits at $95\%$ confidence level (CL) on the signal counts $n_{sig,95}(E_{o})$ in any observed energy bin. The values of $n_{sig,95}(E_{o})$ have been calculated from the observed counts in the on and off regions, $n_{on}(E_{o})$ and $n_{off}(E_{o})$, implementing the Bayesian method in ref.~\cite{loparco2011bayesian} and taking the exposures of the two regions into account. Here we assume that the counts in the off region originate from background, while those in the on region originate from both signal and background, i.e. $n_{off}(E_{o})$ is a Poisson random variable with average value $n_{bkg}(E_{o})$, while $n_{on}(E_{o})$ is a Poisson random variable with average value $n_{sig}(E_{o})+c ~n_{bkg}(E_{o})$, where $c$ is a constant which takes into account the exposures of the two regions. The Bayesian method allows the evaluation of the posterior probability distribution function (PDF) of the signal counts starting from the observed counts $n_{on}(E_{o})$ and  $n_{off}(E_{o})$ and assuming uniform prior PDFs for both $n_{bkg}(E_{o})$ and $n_{sig}(E_{o})$.
Fig.~\ref{fig:calclimits} shows an example of application of this approach to evaluate the upper limits at $95\%$ CL on the DM gamma-ray fluxes in the scenario of the mediator decay channel $\phi \rightarrow b\bar{b}$, assuming $m_{\phi}=11.5\unit{GeV~c^{-2}}$.
We have also evaluated $n_{sig,95}(E_{o})$ with the frequentist method~\cite{Workman:2022ynf,Cranmer:2005hi}, finding no significant differences with respect to the calculation performed with the Bayesian method. 

On the other hand, in the optimistic approach we evaluate $k_{DM}$ assuming a saturated model~\cite{Workman:2022ynf} for the counts in the on region. In this approach we have implemented a hypothesis test based on the maximum likelihood formalism. In the null hypothesis we assume that the observed counts $n_{on}(E_{o})$ in each energy bin are originated from Poisson distributions with average values $n_{on}(E_{o})$, while in the alternative hypothesis we assume that the observed counts $n_{on}(E_{o})$ are originated from Poisson distributions with average values $n_{on}(E_{o})+k_{DM} \cdot \mu_{DM,0}(E_{o})$, where $k_{DM}$ is the DM normalization constant and $\mu_{DM,0}(E_{o})$ are the expected DM counts when the cross section assumes its reference value $\sigma_{0}$. We then define the log-likelihood ratio as:

\begin{equation}
    \lambda(k_{DM}) = \sum_{E_{o}}{ \left[ -k_{DM}\cdot \mu_{DM,0}(E_{o}) - n_{on}(E_{o}) \log \left(
    \frac{n_{on}(E_{o}) +k_{DM} \cdot \mu_{DM,0}(E_{o})}{n_{on}(E_{o})}\right) \right]}.
\label{eq:like-ratio}
\end{equation}
The best fit of $k_{DM}$ is obtained by maximizing the log-likelihod ratio in eq.~\ref{eq:like-ratio}. The upper limit at $95\%$ CL on the normalization constant $k_{DM}$ is evaluated by solving the equation $\lambda = \lambda_{max} - 2.71/2$, where $\lambda_{max}$ is the maximum value of the log-likelihood ratio.

The limits on $k_{DM}$ obtained with the two approaches can be converted into constraints on the capture rate $\Gamma_{\text{cap}}$ and then on the DM-nucleon spin-dependent and spin-independent scattering cross sections. In fact, from eqs.~\ref{eq:generalphidm}, ~\ref{eq:phidm2} and ~\ref{eq:DMyieldFit}, it follows that: 

\begin{equation}
\Gamma_{\text{cap},95\%} = k_{DM,95\%}\cdot 4\pi D^2
\label{eq:caplim}
\end{equation}
where we have indicated with $\Gamma_{\text{cap},95\%}(m_{\chi},\sigma)$ the upper limit at $95\%$ CL on the capture rate and with $k_{DM,95\%}$ the corresponding upper limit on the normalization $k_{DM}$. The constraints on the capture rate $\Gamma_{\text{cap}}$ are model-independent, as no hypothesis is required on the DM-nucleon scattering process, which can be either spin-dependent or spin-independent. However, in writing eq.~\ref{eq:generalphidm}, it is implicitly assumed that the DM capture rate $\Gamma_{cap}$ is equal to the annihilation rate $\Gamma_{ann}$, i.e. an equilibrium scenario between DM capture and annihilation is implicitly assumed. 

\begin{figure*}[t]
\centering
\includegraphics[width=0.98\columnwidth]{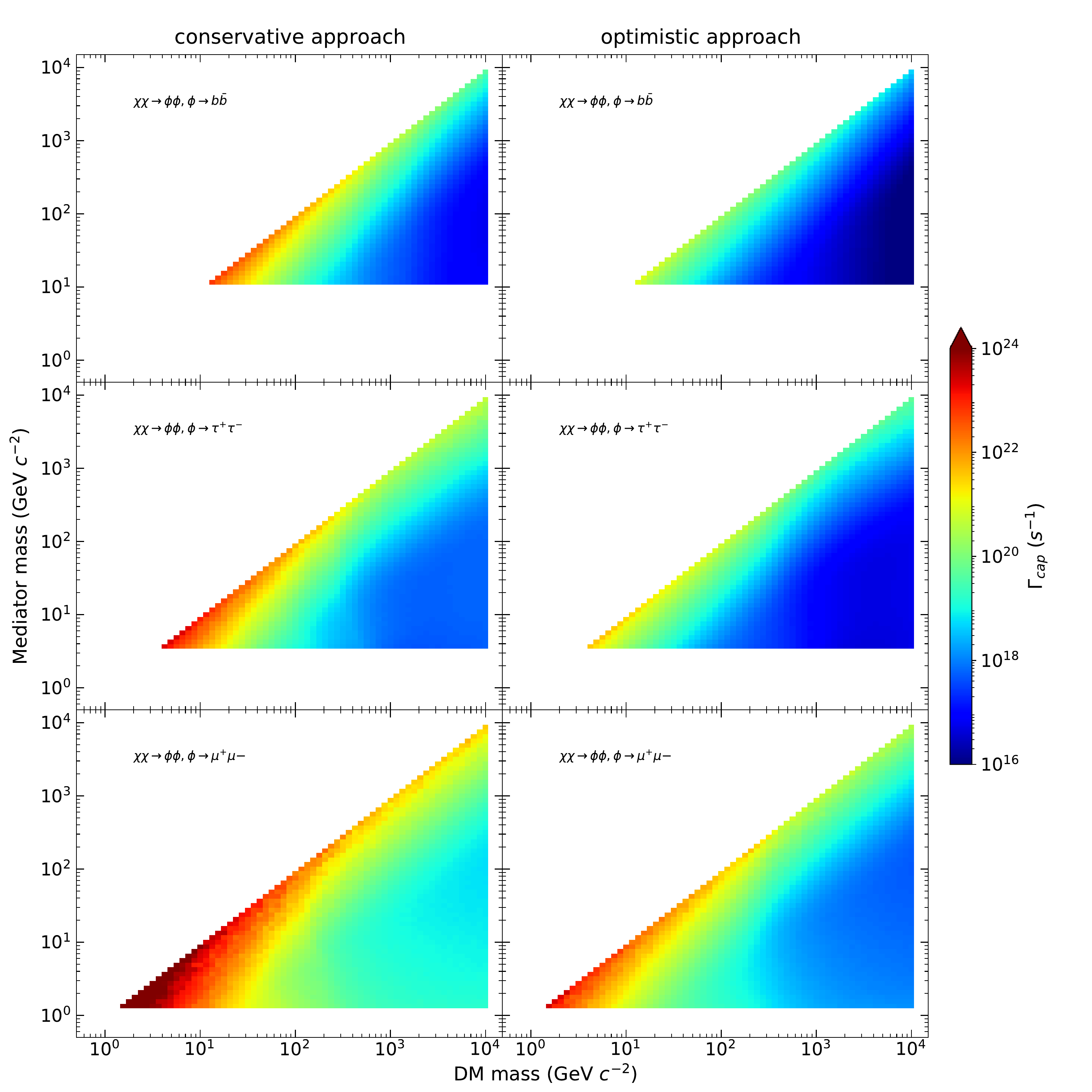}
\caption{Upper limits at $95\%$ CL for the capture rate $\Gamma_{\text{cap}}$ as a function of the DM and mediator masses evaluated with the conservative (left plots) and with the optimistic approaches (right plots). The plots have been obtained under the assumptions that the mediator decays into the $b\bar{b}$ (top panels), $\tau^{+}\tau^{-}$ (middle panels) and $\mu^{+}\mu^{-}$ (bottom panels) channels.}
\label{fig:CapRateLims}
\end{figure*}

Fig.~\ref{fig:CapRateLims} summarizes the limits on the capture rate as a function of the DM and mediator masses $m_{\chi}$ and $m_{\phi}$, evaluated in the mediator decay channels $b\bar{b}$, $\tau^{+}\tau^{-}$ and $\mu^{+}\mu^{-}$ with the two analysis approaches implemented in this work. 

Finally, to turn the constraints on the capture rate into constraints on the DM-nucleon scattering cross sections (for both the spin-dependent and spin-independent cases), we have used the results of ref.~\cite{cuoco2020search}, where we evaluated the capture rates $\Gamma_{\text{cap}}(m_{\chi},\sigma_{0})$ at the reference cross section $\sigma_{0}=10^{-40}\unit{cm^{2}}$. Since the capture rate is proportional to the scattering cross section, the upper limit at $95\%$ C.L. on the cross section is given by:

\begin{equation}
\sigma_{95\%} = \sigma_{0}~ \frac{\Gamma_{\text{cap},95\%}}{\Gamma_{\text{cap}}(m_\chi,\sigma_0)}
\label{eq:sigmalim}
\end{equation}

\section{Results and discussion}
\label{sec:results}

\begin{figure*}[t]
\centering
\includegraphics[width=0.98\columnwidth]{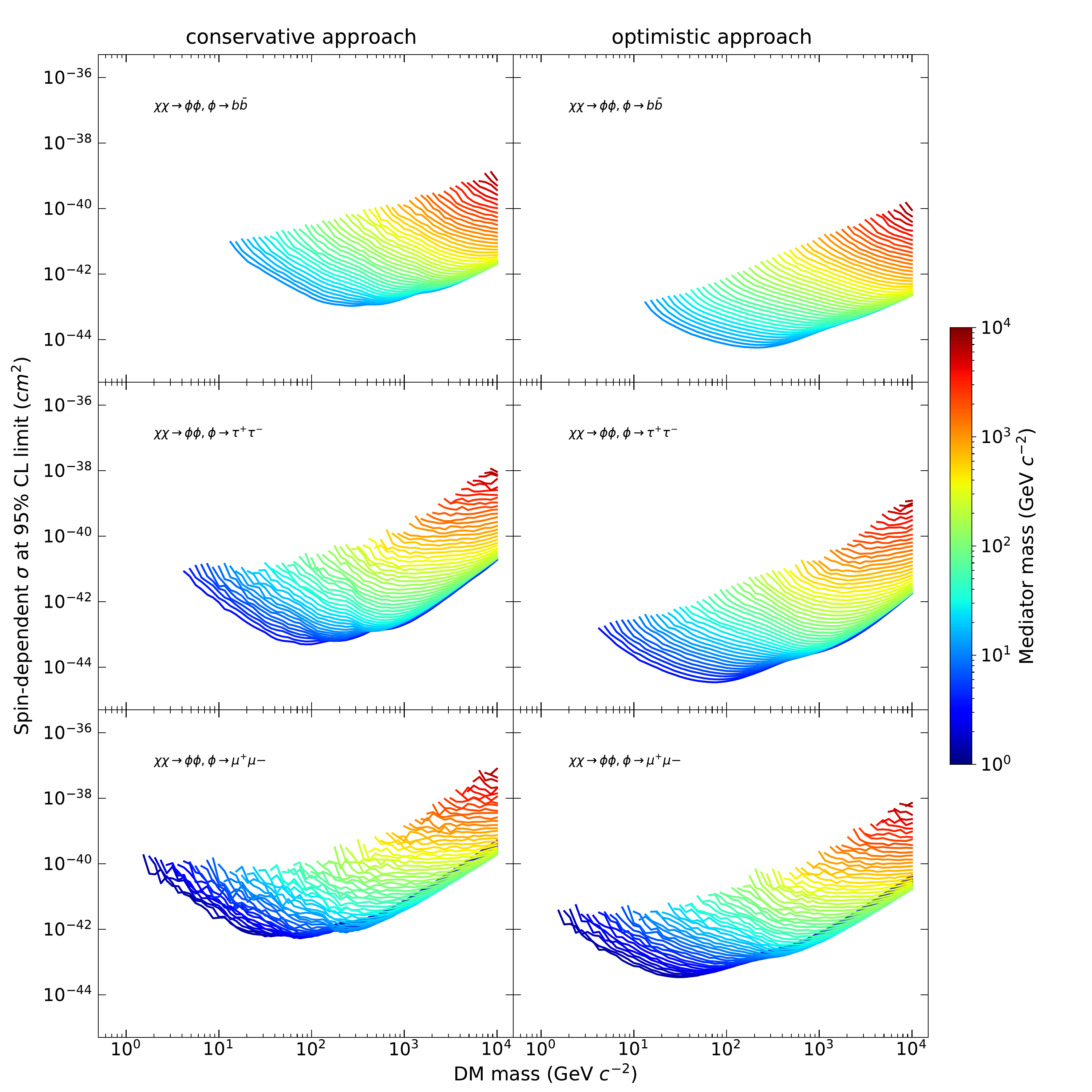}
\caption{Upper limits at $95\%$ CL on the DM-nucleon scattering spin-dependent cross section $\sigma_{SD}$ as a function of $m_{\chi}$. The limits have been obtained under the assumptions of DM annihilation into long-lived mediators decaying in the $b\bar{b}$ (top panels), $\tau^{+}\tau^{-}$ (middle panels) and $\mu^{+}\mu^{-}$ (bottom panels) channels, with a decay length $L=R_{\odot}$. The colored lines correspond to different values of the mediator mass $m_\phi$. The plots in the left column show the results obtained with the conservative approach, while those in the right column show the results obtained with the optimistic approach.}
\label{fig:sigmaSD}
\end{figure*}

\begin{figure*}[t]
\centering
\includegraphics[width=0.98\columnwidth]{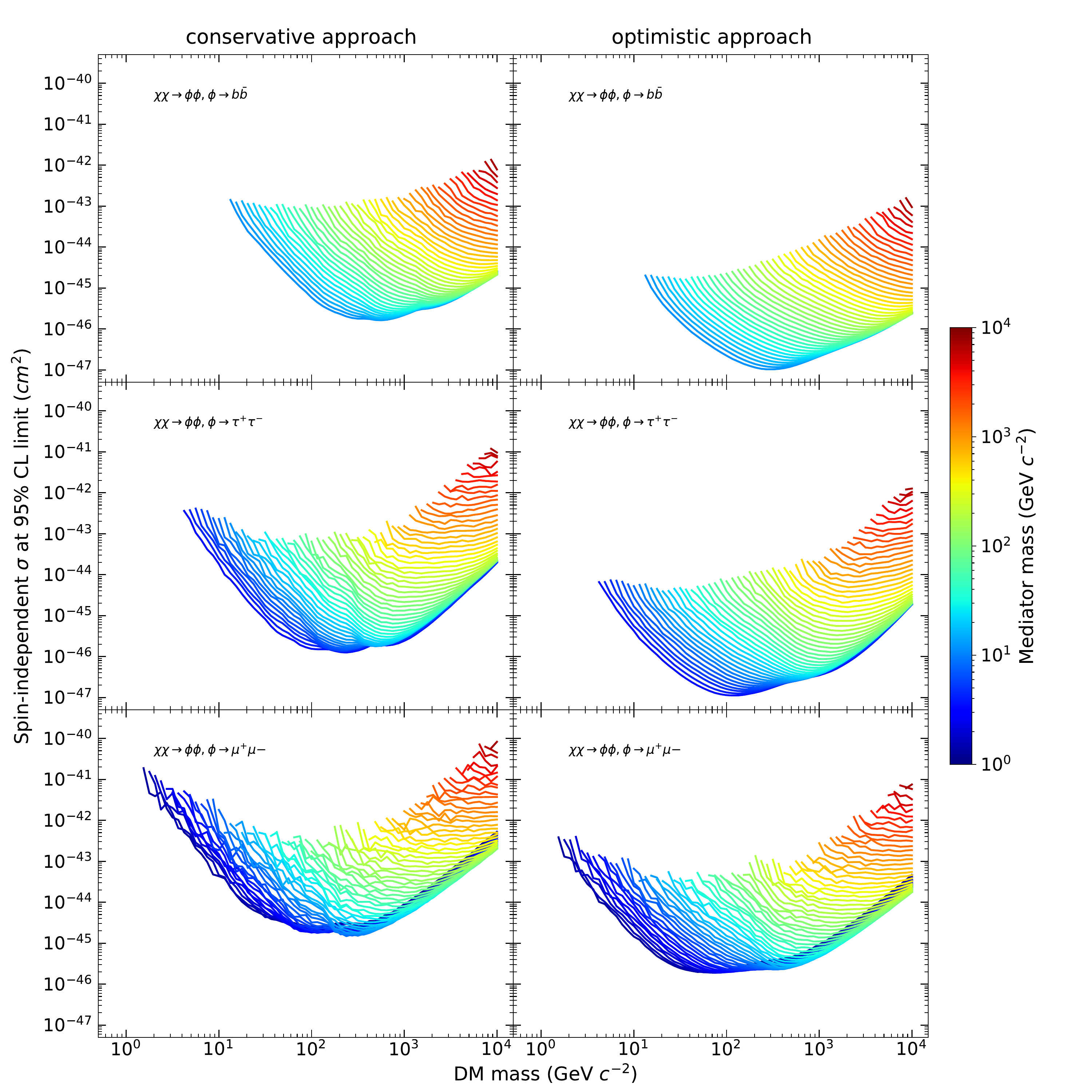}
\caption{Upper limits at $95\%$ CL on the DM-nucleon scattering spin-independent cross section $\sigma_{SI}$ as a function of $m_{\chi}$. The limits have been obtained assuming DM annihilation into long-lived mediators decaying in the $b\bar{b}$ (top panels), $\tau^{+}\tau^{-}$ (middle panels) and $\mu^{+}\mu^{-}$ (bottom panels) channels, with a decay length $L=R_{\odot}$.  The colored lines correspond to different values of the mediator mass $m_\phi$. The plots in the left column show the results obtained with the conservative approach, while those in the right column show the results obtained with the optimistic approach.}
\label{fig:sigmaSI}
\end{figure*} 

In Figs.~\ref{fig:sigmaSD} and \ref{fig:sigmaSI} the upper limits at $95\%$ CL on the spin-dependent ($\sigma_{SD}$) and spin-independent ($\sigma_{SI}$) DM-nucleon cross sections are shown as functions of the mass $m_{\chi}$ of the DM particles and for the mediator decay channels $\phi \rightarrow b\bar{b}$, $\phi \rightarrow \tau^{+}\tau^{-}$ and $\phi \rightarrow \mu^{+}\mu^{-}$. These constraints have been evaluated for a set of  mediator masses $m_{\phi}$ in the kinematically allowed range, under the assumptions of a mediator decay length $L=R_{\odot}$. In the plots we show the limits evaluated with the two analysis approaches discussed in sec.~\ref{sec:analysis}. As expected, we see that the constraints on the cross sections obtained with the optimistic approach are in general stronger than those obtained with the conservative approach. The differences between the limits obtained with the two analysis approaches are typically within two orders of magnitude.

For all the channels investigated in this analysis the limits on the cross sections $\sigma_{SD}$ and $\sigma_{SI}$ are in the ranges from $10^{-45}$ up to $10^{-39}$ $\unit{cm^2}$ and from $10^{-47}$ up to $10^{-42}$ $\unit{cm^2}$ respectively. The strongest constraints are obtained for the lowest allowed values of $m_\phi$, i.e. when the mediator is highly boosted ($m_\phi \ll m_\chi$). 
We remark here that the limits on the DM-nucleon scattering cross section depend on the calculation of the reference capture rate. As discussed in Sec.~\ref{sec:medchannelstheory}, for this calculation we have used the \texttt{DARKSUSY} code with the default settings for the parameters describing the local DM density and its velocity distribution. Any variations of these parameters would result in a change of the reference capture rate, and consequently of the limits on the scattering cross sections. The values of the default parameters of \texttt{DARKSUSY} are widely used in the literature, and a study of the variations of the capture rate with these parameters would go beyond the goals of the present work.

As mentioned above, in our analysis we have considered DM masses in the kinematically allowed range. However, for masses below a few \unit{GeV}, the process of DM evaporation in the Sun interior can be non-negligible~\cite{Griest:1986yu}, and should be taken into account in the evaluation of the limits.

\begin{figure*}[t] 
\centering
    \includegraphics[width=0.48\columnwidth,height=0.25\textheight]{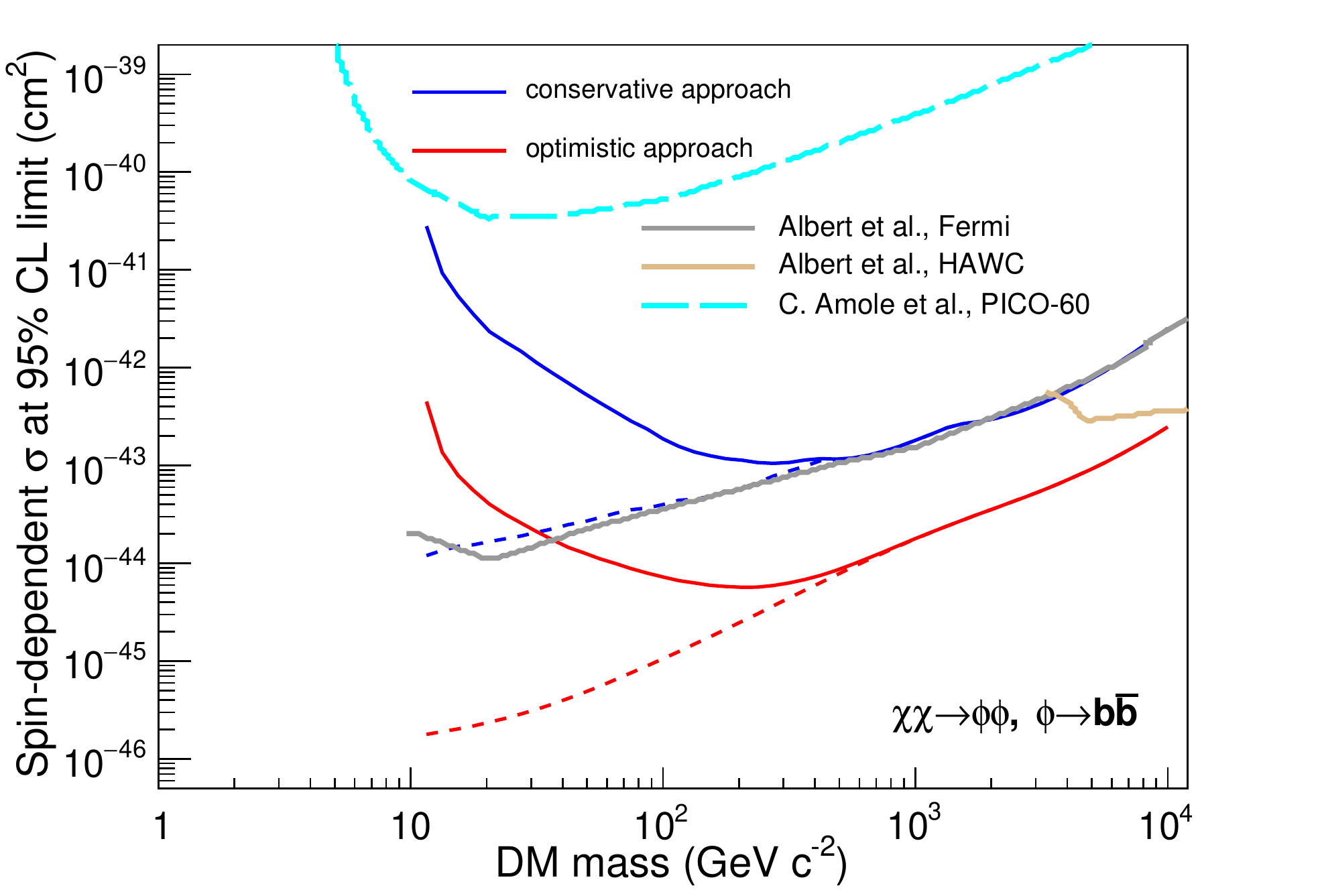}
    \includegraphics[width=0.48\columnwidth,,height=0.25\textheight]{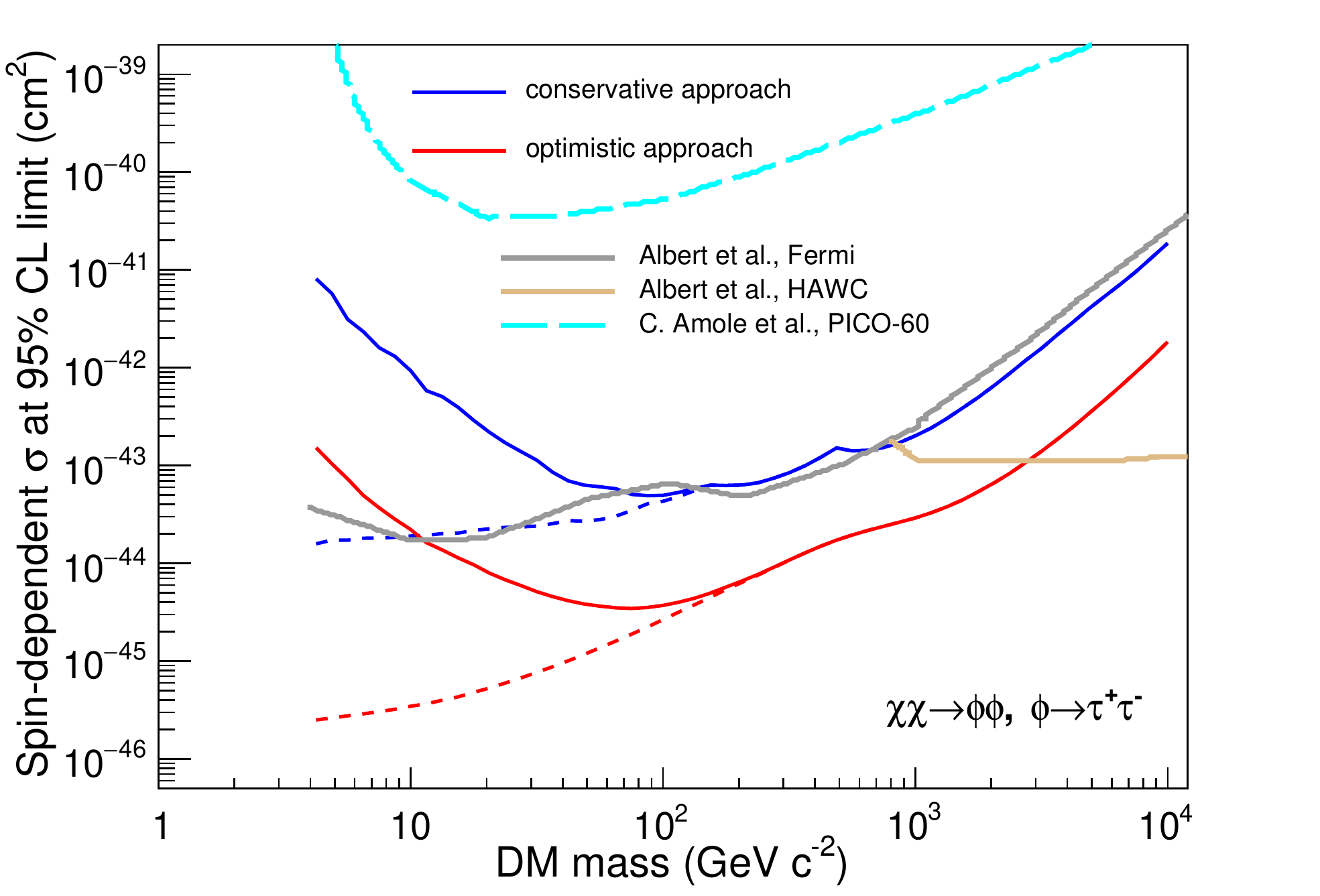}  
    \includegraphics[width=0.48\columnwidth,height=0.25\textheight]{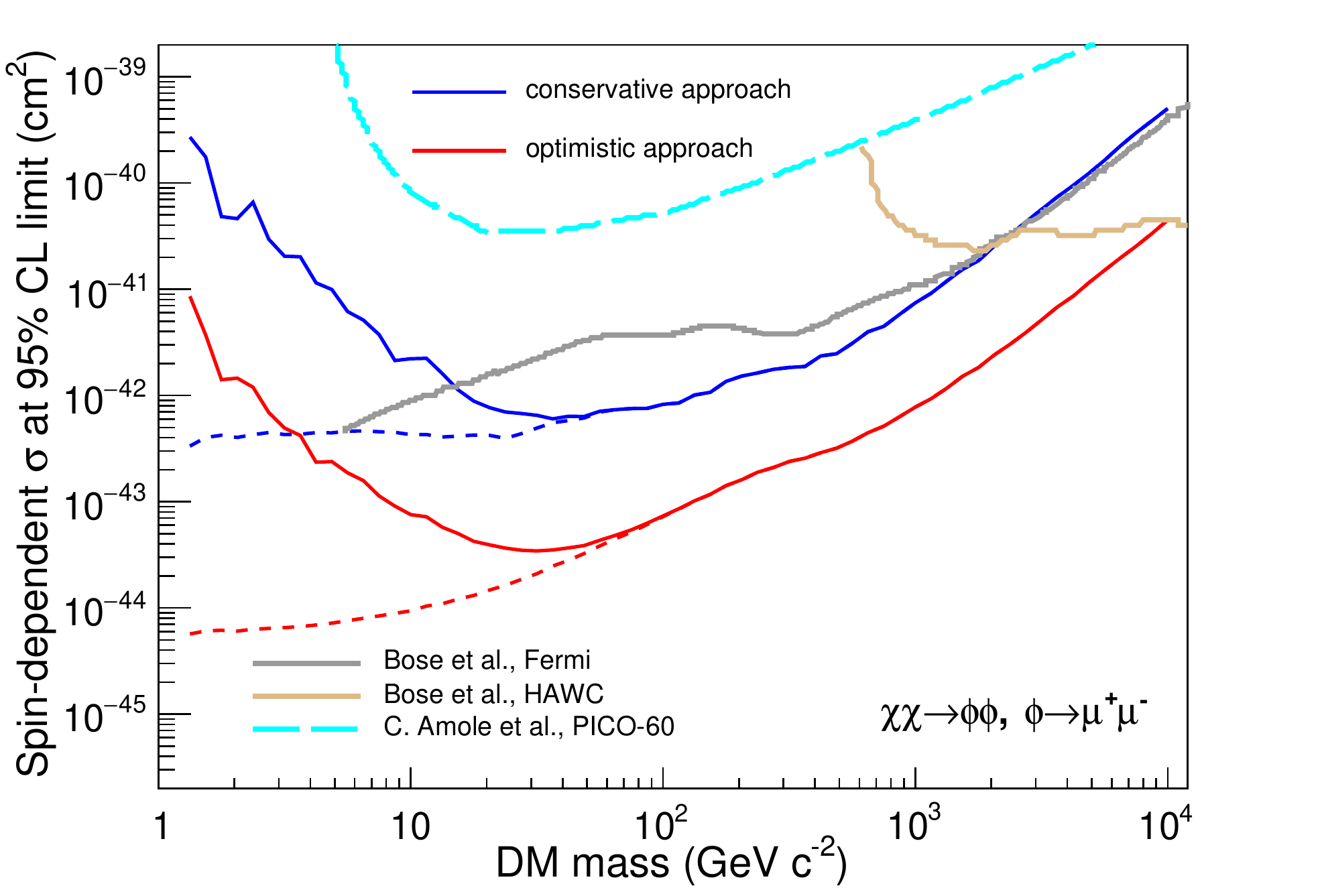}  
    \includegraphics[width=0.48\columnwidth,height=0.25\textheight]{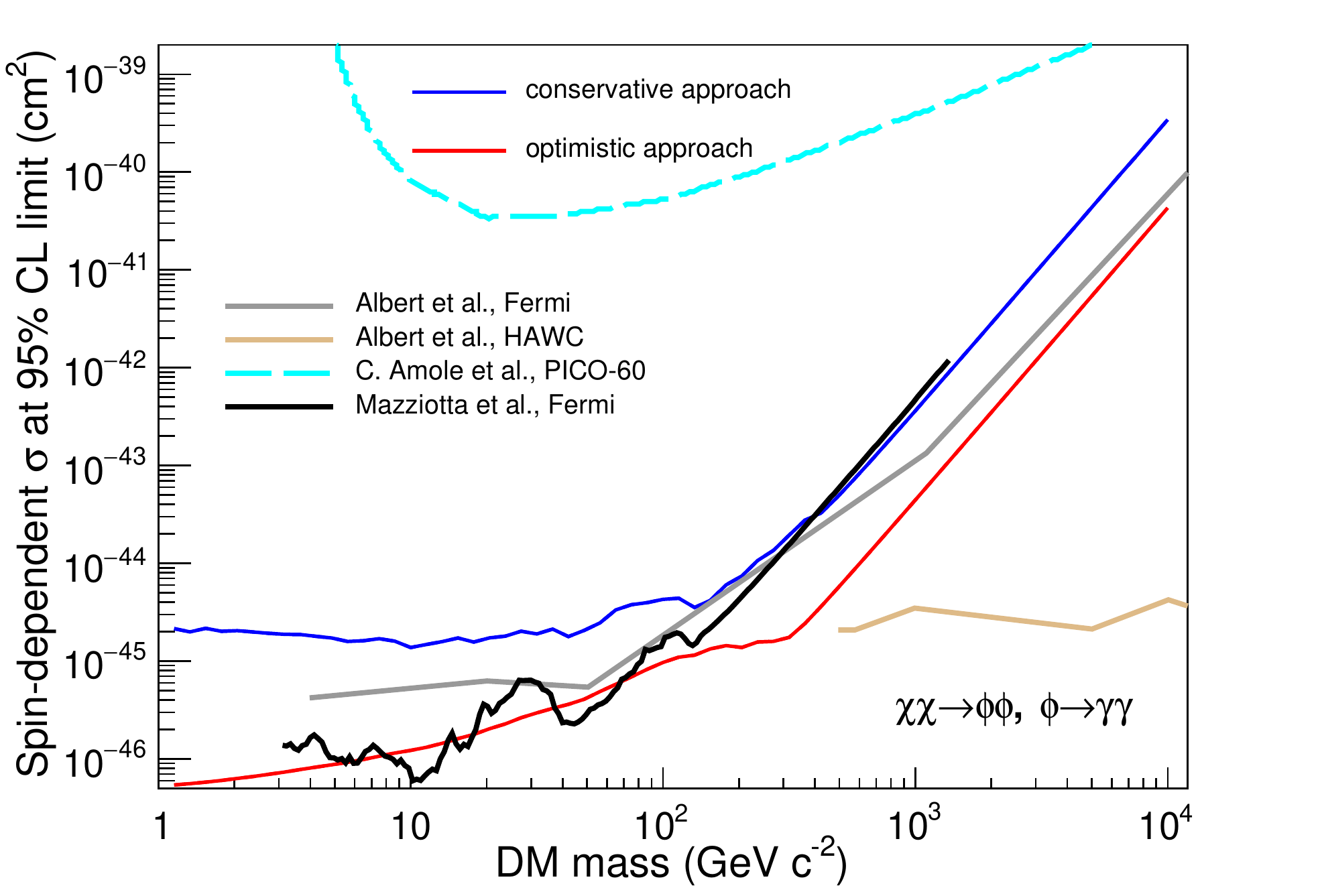} 
\caption{Upper limits at $95\%$ CL on the spin-dependent DM-nucleon scattering cross section $\sigma_{SD}$ for the scenario with DM particles annihilating into long-lived mediators, which can decay in the $b\bar{b}$ (top-left panel), $\tau^+\tau^-$ (top-right panel), $\mu^+\mu^-$ (bottom-left panel) and $\gamma\gamma$ (bottom-right panel) channels, with decay length $L=R_{\odot}$. For each channel, the constraints have been evaluated assuming for $m_{\phi}$ the minimum allowed value. The blue lines indicate the constraints obtained with the conservative analysis approach, while the red lines indicate those obtained with the optimistic approach. The dashed lines indicate the results obtained when the calculation of the gamma-ray flux from DM is performed without requiring a maximum angular separation of $2^\circ$ of gamma rays from the direction of the Sun. The results obtained in this work are compared with those from HAWC and Fermi (refs.~\cite{albert2018constraints,Bose:2021cou}) within the same theoretical scenario. The $90\%$ CL limits obtained from direct measurements of the spin-dependent DM-nucleon cross sections performed by the PICO-60 experiment~\cite{Amole:2019fdf} are also shown. In the case of the $\gamma \gamma$ channel we also show the results of our previous analysis with 10 years of LAT data~\cite{mazziotta2020search}.}
\label{fig:summary1}
\end{figure*}

The first three panels of Fig.~\ref{fig:summary1} show the upper limits obtained in this work on the spin-dependent DM-nucleon scattering cross section $\sigma_{SD}$ in the scenarios in which the mediator can decay in the channels $b\bar{b}$, $\tau^{+}\tau^{-}$ and $\mu^{+}\mu^{-}$. The limits shown in Fig.~\ref{fig:summary1} have been evaluated assuming a mediator decay length $L=R_{\odot}$ and choosing for $m_{\phi}$ the lowest kinematically allowed value in each channel. We have also repeated our analysis procedure using the gamma-ray fluxes evaluated using \texttt{med\_dec} without requiring the maximum angular separation of $2\degrees$ between the arrival direction of gamma rays at Earth and the direction of the Sun (dashed lines in the figure). We find that if the DM fluxes are evaluated without any cut on the arrival directions of photons at Earth, the constraints on the cross sections are stronger. This feature is clearly evident in the region of low DM masses ($m_{\chi}<1\unit{TeV}$) and is due to the fact that, as discussed in Sec.~\ref{sec:medchannelstheory}, the cut on the arrival directions reduces the intensity of the expected DM gamma-ray signal in the low-energy region. Since for low DM masses the limits on the DM signal intensity are basically set by the limits on the gamma-ray counts in the low-energy bins ($E < 10\unit{GeV}$), a lower number of expected counts from DM gamma rays in those bins will result in an increased limit on the DM signal intensity and consequently on the cross section. 

In Fig.~\ref{fig:summary1} we also compare the limits on $\sigma_{SD}$ obtained in our analysis with those published in refs.~\cite{albert2018constraints,Bose:2021cou}, obtained from other analyses of the Fermi LAT and of the HAWC data, and with the constraints at $90\%$ CL from the direct measurements performed by {the PICO-60 experiment}~\cite{Amole:2019fdf}. Our limits are comparable to other measurements for a DM mass range up to $\sim 1 \unit{TeV}$. 

Finally, we have also reconsidered the scenario which we studied in our previous work~\cite{mazziotta2020search}, with the mediator decaying directly into pairs of gamma rays. The last panel of Fig.~\ref{fig:summary1} shows the upper limits on $\sigma_{SD}$ evaluated in this scenario compared with those published in ref.~\cite{mazziotta2020search}. 
The limits obtained with the optimistic analysis approach are consistent with those of our previous work up to DM masses of $\sim 100 \unit{GeV/c^{2}}$, while they become consistent with those obtained with the conservative approach for DM masses in the $\unit{TeV}$ range.

\begin{figure*}[t]
\centering

\includegraphics[width=0.48\columnwidth,height=0.25\textheight]{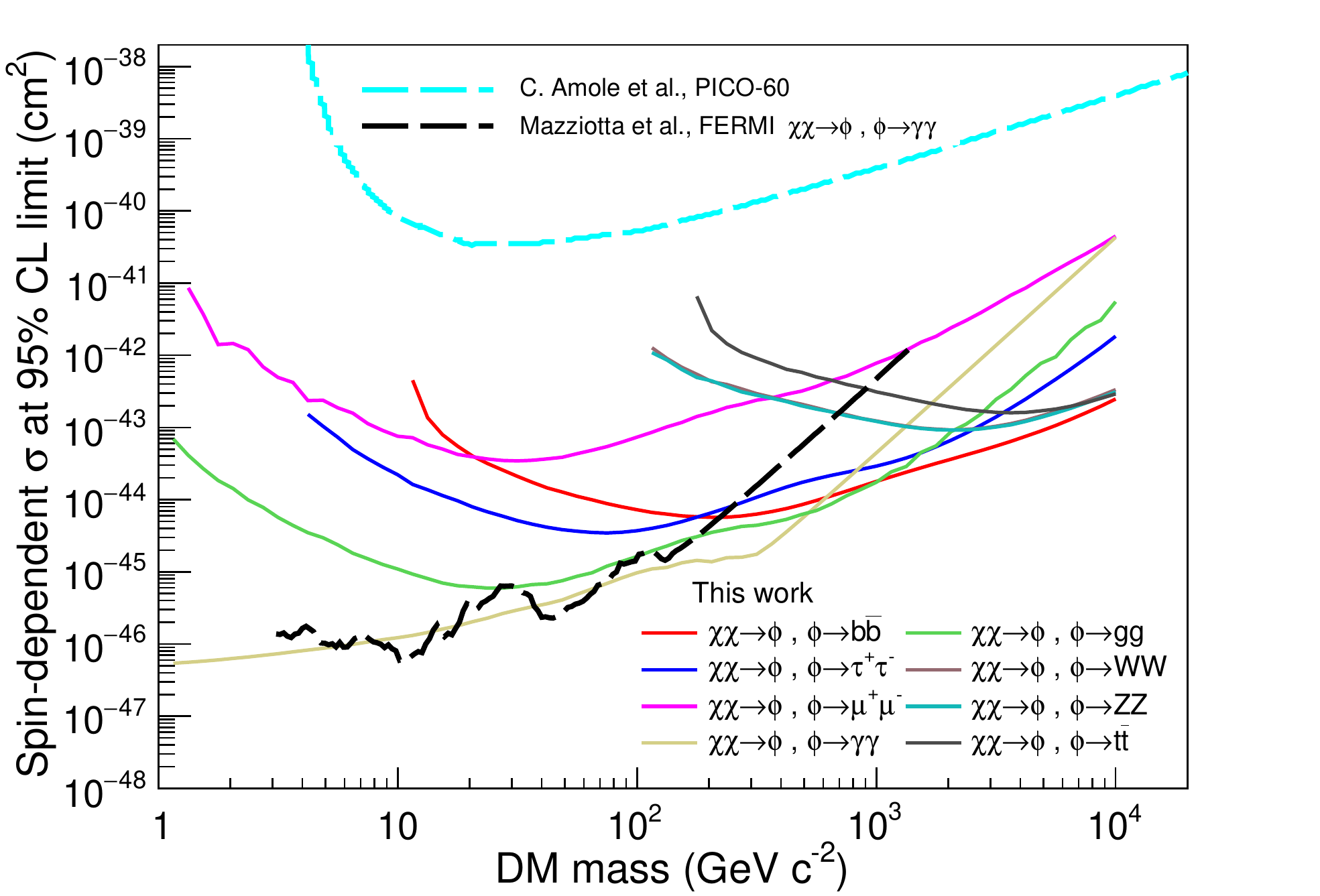}
\includegraphics[width=0.48\columnwidth,height=0.25\textheight]{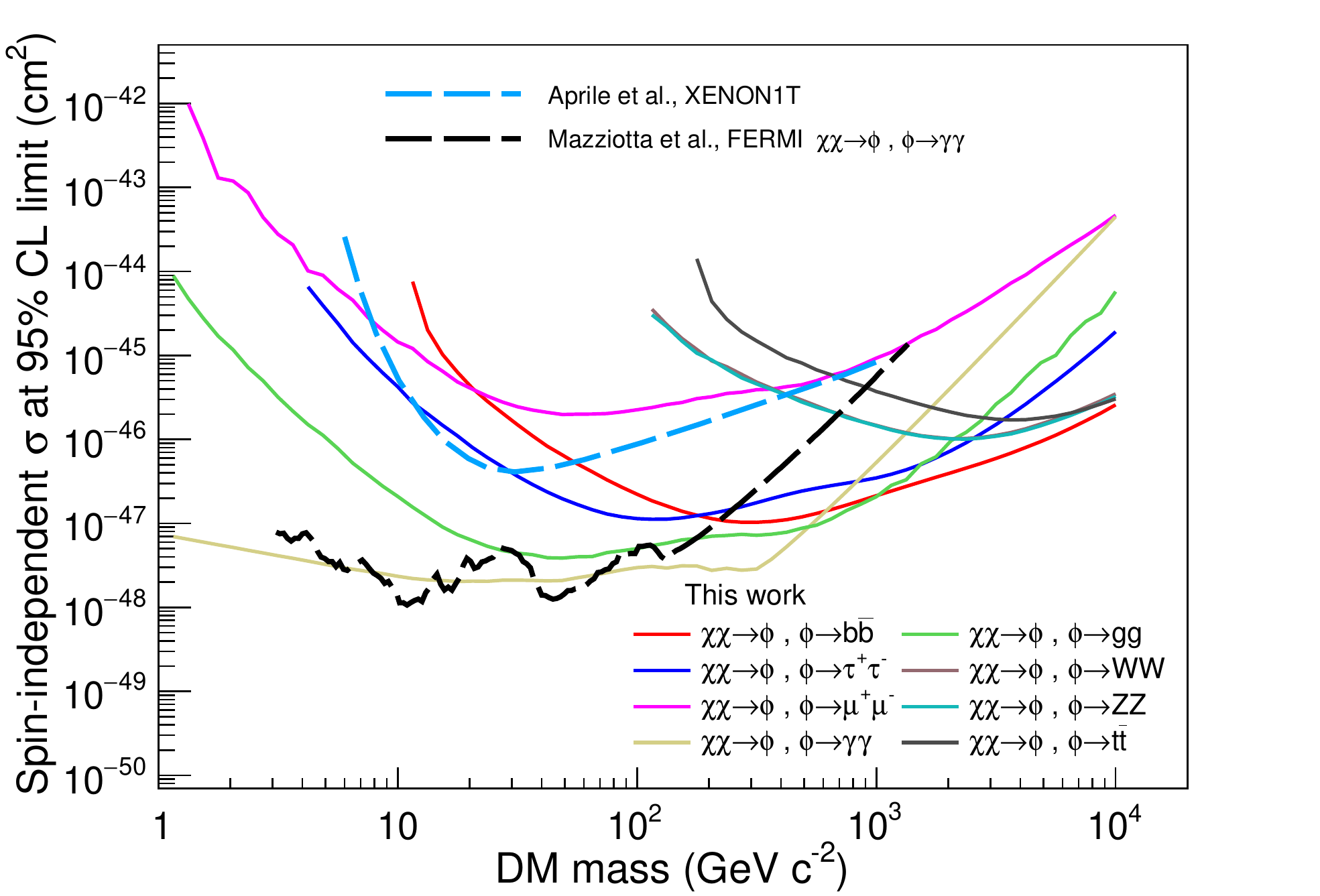}
\caption{Upper limits at $95\%$ CL on the spin-dependent (left) and spin-independent (right) DM-nucleon scattering cross section for the scenario with DM particles annihilating into long-lived mediators, which can decay in the $b\bar{b}$, $\tau^+\tau^-$, $\mu^+\mu^-$, $\gamma\gamma$, $gg$, $WW$, $ZZ$ and $t\bar{t}$ channels, with decay length $L=R_{\odot}$. The constraints have been evaluated with the optimistic analysis approach. For each channel we have assumed the minimum kinematically allowed value for $m_{\phi}$. The $90\%CL$ limits obtained from the direct measurements of the spin-dependent cross sections performed by the PICO-60 experiment~\cite{Amole:2019fdf} and of the spin-independent cross sections performed by the XENON1T experiment~\cite{Aprile:2018dbl} are also shown.}
\label{fig:summary2}
\end{figure*}

In Fig.~\ref{fig:summary2} we summarize the constraints obtained in this work with the optimistic analysis approach for both the spin-dependent and spin-independent DM-nucleon scattering cross sections $\sigma_{SD}$ and $\sigma_{SI}$. The limits shown in the figure have been obtained by studying several possible mediator decay channels ($b\bar{b}$, $\tau^+\tau^-$, $\mu^+\mu^-$, $\gamma\gamma$, $gg$, $WW$, $ZZ$ and $t\bar{t}$). We have assumed a mediator decay length $L=R_{\odot}$ and for each channel we have plotted the limits obtained by selecting the lowest kinematically allowed value of $m_{\phi}$. Our results are compared with those obtained from the direct measurements performed by the XENON1T~\cite{Aprile:2018dbl} and PICO-60~\cite{Amole:2019fdf} experiments respectively. In the case of the spin-dependent cross section, the limits obtained in this work are a few orders of magnitude stronger than those obtained by PICO-60 in the whole DM mass range explored, while in the case of the spin-independent cross section the limits are consistent with those from XENON1T. We also remark that the constraints quoted in refs.~\cite{Aprile:2018dbl} and~\cite{Amole:2019fdf} are upper bounds at $90\%$ CL, while here $95\%$ CL limits are presented.    

The evaluation of the constraints on the DM-nucleon cross section is strongly model-dependent, since the mediator properties determine the final results. However, we point out that the model considered in this work is the same investigated by other authors in their recent works~\cite{albert2018constraints,Leane:2017vag,cuoco2020search,Bose:2021cou}. In particular, in Ref.~\cite{albert2018constraints}, the authors evaluated the limits on $\sigma_{SD}$ in a scenario with the mediator decaying into the $b\bar{b}$,  $\tau^+\tau^-$ and $\gamma\gamma$ channels with a decay length $L=R_{\odot}$, combining Fermi and HAWC 3-years observations. The same scenario was investigated also in Ref.~\cite{Bose:2021cou} for the $\mu^+\mu^-$ decay channel. The final limits quoted in these works are those obtained with the lowest allowed mediator mass, since, as already discussed, this condition provides the strongest limits.

Finally, we point out here that the limits presented in this work have been evaluated under the assumption of equilibrium between the DM capture and annihilation processes in the solar environment and assuming $100\%$ branching ratios for the processes $\chi \chi \rightarrow \phi \phi$ and $\phi \rightarrow XX$. 

The results obtained in this work show the potentiality of solar gamma rays as a probe for indirect DM detection, since the limits obtained with this analysis are comparable or even stronger than those currently quoted in the literature. We also remark that searches from solar DM could benefit from an accurate model of the steady solar emission, that could improve the description of the background.

\begin{acknowledgments} 
The Fermi LAT Collaboration acknowledges generous ongoing support from a number of agencies and institutes that have supported both the development and the operation of the LAT as well as scientific data analysis. These include the National Aeronautics and Space Administration and the Department of Energy in the United States, the Commissariat \`a l'Energie Atomique and the Centre National de la Recherche Scientifique / Institut National de Physique Nucl\'eaire et de Physique des Particules in France, the Agenzia Spaziale Italiana  and the Istituto Nazionale di Fisica Nucleare in Italy, the Ministry of Education, Culture, Sports, Science and Technology (MEXT), High Energy Accelerator Research Organization (KEK) and Japan Aerospace Exploration Agency (JAXA) in Japan, and the K.~A.~Wallenberg Foundation, the Swedish Research Council and the Swedish National Space Board in Sweden.
 
Additional support for science analysis during the operations phase is gratefully acknowledged from the Istituto Nazionale di Astrofisica in Italy and the Centre National d'\'Etudes Spatiales in France. This work performed in part under DOE Contract DE-AC02-76SF00515.
\end{acknowledgments}

\bibliographystyle{unsrt}
\bibliography{DMSunMediatorLAT_2022.bib}{}
\end{document}